\documentclass{moriond}

\usepackage{amssymb}
\usepackage{xspace}

\def\be{\begin{equation}}
\def\ee{\end{equation}}
\def\bea{\begin{eqnarray}}
\def\eea{\end{eqnarray}}

\newcommand{\sqrts}{\sqrt{\rm s}}

\newcommand{\epem}{e^+e^-}

\providecommand{\ffbar}{f\overline{f}}
\providecommand{\bbbar}{b\overline{b}}

\newcommand{\AFBb}  {A_{_{\textsc{fb}}}^{0,b}}

\newcommand{\AFBobs}  {A_{_{\textsc{fb}}}^{obs,b}}
\newcommand{\weakang}  {\sin^2\theta_{\rm W}}

\newcommand{\pythia}{{\sc pythia}}
\newcommand{\vincia}{{\sc vincia}}
\newcommand{\jetset}{{\sc jetset}}

\def\mean#1{\ensuremath{\left<#1\right>}}

\newcommand{\Photo}{\includegraphics[height=30mm]{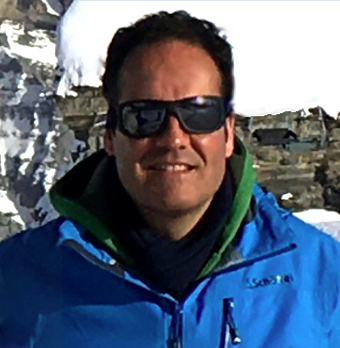}}

\begin{document}
\title{Forward-backward $b$-quark asymmetry at the Z pole: QCD uncertainties redux}


\author{\underline{David d'Enterria}$^1$, and Cynthia Yan$^{1,2}$}
\address{$^1$ CERN, EP Department, CH-1211 Geneva 23, Switzerland\\
$^2$ Harvey Mudd College, Department of Physics, Claremont CA91711, USA}

\maketitle\abstracts{
The forward-backward asymmetry of $b$-quarks measured at LEP in $\epem$ collisions at the Z pole, 
$\AFBb|^{\rm exp} = 0.0992\pm0.0016$, remains today the electroweak precision observable with the largest 
disagreement (2.8$\sigma$) with the Standard Model theoretical 
prediction, $\AFBb|^{\rm th} = 0.1037\pm0.0008$. The dominant systematic uncertainties 
are due to QCD effects --- $b,c$-quark showering and fragmentation, and $B,D$ meson 
decay models --- which have not been revisited in the last 20 years. We reassess the QCD uncertainties
of the eight original LEP measurements of $\AFBb$, using modern parton shower simulations
based on \pythia~8 and \pythia\,8\,$+$\,\vincia\ with different tunes of soft and collinear radiation 
as well as of hadronization. Our analysis indicates QCD uncertainties, of order $\pm$0.4\% and $\pm$1\%
for the jet-charge and lepton-charge based analyses, that are overall slightly smaller but still 
consistent with the original ones. Using the updated QCD systematic uncertainties, we obtain $\AFBb = 0.0996\pm0.0016$.
}

\section{Introduction}

In the Standard Model (SM), the Z boson mediates weak neutral currents between fermions of
the same generation. The Z couples to both left- and right-handed chiral states with different 
strengths depending on weak-isospin and electromagnetic charges. The vector 
and axial-vector Z couplings for a fermion of type $f$ are $g^f_V = (g_L^f + g_R^f) = I_3^f - 2Q^f\weakang$ 
and $g^f_A = (g_L^f -g_R^f) = I_3^f$ respectively, where $I_3$ is the third component of the weak isospin of the fermion, 
$Q^f$ its charge (related to the former via the hypercharge $Y^f$: $Q^f=I_3^f+Y^f/2$), and $\weakang\approx 0.23$ 
is the weak mixing angle 
that controls the $\gamma$--Z mixing and provides a relationship between
the coupling constants of the electroweak theory: $g\sin\theta_W=g'\cos\theta_W=e$.
From the expressions above, the varying strengths of the Z-fermion couplings for the
$(\nu_e,\nu_\mu,\nu_\tau)$, $(e,\mu,\tau)$, $(u,c,t)$, and $(d,s,b)$ lepton/quark groups are explained.
The mixed Z vector and axial-vector couplings not only affect the total $\epem\to\ffbar$ cross section but
induce asymmetries in the angular distributions of the final-state fermions produced 
in the process. 
Angular asymmetries in the $\epem\to\ffbar$ final-state are ultimately driven by the fermions' charge $Q$ and the weak mixing angle:
\begin{eqnarray}
\mathcal{A}_f=\frac{(g_{L}^f)^2-(g_{R}^f)^2}{(g_{L}^f)^2+(g_{R}^f)^2}=2\frac{g_{V}^f/g_{A}^f}{1+(g_{V}^f/g_{A}^f)^2}\,,
\mbox{ with }\;\;\frac{g_{V}^f}{g_{A}^f}=1-4|Q_f|\sin^2\theta^f_{\mbox{eff}}\,.
\end{eqnarray}
Experimentally, forward-backward asymmetries in $\epem\to\ffbar$ are determined from the ratio of the number of forward- (backward-)going
(anti)fermions measured in the hemisphere defined by the direction of the $e^+$ ($e^-$) beams: 
\begin{eqnarray}
A_{\rm FB}^f=\frac{N_F-N_B}{N_F+N_B},\;\;\mbox{ where }\;F=\int_0^1 \frac{d\sigma}{d\Omega}d\Omega,\;\;B=\int_{-1}^0 \frac{d\sigma}{d\Omega}d\Omega,\;
\end{eqnarray}
The forward-backward asymmetry of $b$ quarks ($\AFBb$) in the process $\epem\to Z \to\bbbar$ at $\sqrts = m_Z$
is the one most accurately measured among all quarks at LEP, given that $b$-quarks are the easiest jets to identify. The 
value $\AFBb|^{\rm exp} = 0.0992\pm0.0016$, obtained from the combination of eight measurements at $\sqrts = 91.21$--91.26~GeV 
using two different (lepton- and jet-charge based) methods, shows today the largest discrepancy ($2.8\sigma$) 
with respect to the theoretical SM prediction, $\AFBb|^{\rm th} = 0.1037\pm0.0008$
(and so does the value of $\weakang$ derived from them)~\cite{ALEPH:2005ab}. 
We reanalyze here the original studies to see if such a discrepancy could be explained by a potential 
underestimation of the associated systematic uncertainties.


\section{LEP $b$-quark forward-backward asymmetry data}
\label{sec:}

Table~\ref{tab:AFBb} lists the eight $\AFBb$ measurements with the breakdown of their uncertainties.
In four measurements, the original $b$,~$\bar{b}$ quarks are identified from the charge of the leading 
lepton $\ell$ inside each $b$-jet (through the fragmentation $b\to B, b\to c\to D$ and subsequent 
$B,D\to\ell$ decay), whereas in the other four, the $b$ charge is reconstructed from 
the jet constituent particles.
The statistical uncertainties of $\AFBb$ dominate, being twice bigger than the systematic ones, while the 
QCD uncertainties account for about half of the latter (and are assumed to be fully-correlated among measurements). 
The QCD-related biases on $\AFBb$ depend strongly on the experimental selection procedure and are related to: (i) hard gluon radiation, and
(ii) smearing of the $b$-jet (thrust) axis due to $b$ and $(b\to)c$ soft radiation and hadronization,
and subsequent $B$ and $D$ hadron decay models. Whereas the first bias is theoretically well controlled
through next-to-next-to-leading-order perturbative QCD (plus massive $b$-quark) corrections~\cite{Bernreuther:2016ccf}, 
the uncertainties of the latter were estimated using Monte Carlo (MC) parton shower 
simulations~\cite{Abbaneo:1998xt} that have not been revisited in 20 years.
\begin{table}[htbp]
\caption[]{LEP measurements of $\AFBb$  and associated statistical, total systematic, and QCD-systematic uncertainties
(with the newly-computed QCD systematics quoted in parentheses).\label{tab:AFBb}}
\begin{center}
\tabcolsep=1.1mm
\begin{tabular}{lcccc}\hline
Measurement &  $\AFBb$ &  & uncertainties & \\
            &          &  stat. & total syst. & QCD syst. (new)\\\hline
ALEPH lepton (2002)~\cite{leptonALEPH} & $0.1003 \pm 0.0038 \pm 0.0017$ &  $4.1\%$ & $1.7\%$ & $0.6\%\,(0.8\%)$ \\
DELPHI lepton (2004-5)~\cite{leptonDELPHI95} & $0.1025 \pm 0.0051 \pm 0.0024$ & $6.4\%$ & $2.4\%$ & $1.5\%\,(1.3\%)$ \\ 
L3 lepton (1999)~\cite{leptonL392} & $0.1001 \pm 0.0060 \pm 0.0035$ & $6.9\%$ & $3.4\%$ & $1.8\%\,(0.8\%)$\\
OPAL lepton (2003)~\cite{leptonOPAL}& $0.0977 \pm 0.0038 \pm 0.0018$ & $4.3\%$ & $1.5\%$ & $1.1\%\,(1.4\%)$ \\\hline
ALEPH jet-charge (2001)~\cite{jetqALEPH} & $0.1010 \pm 0.0025 \pm 0.0012$ & $2.7\%$ & $1.1\%$ & $0.5\%\,(0.5\%)$ \\
DELPHI jet-charge (2005)~\cite{jetqDELPHI} & $0.0978 \pm 0.0030 \pm 0.0015$ & $3.3\%$ & $1.5\%$& $0.5\%\,(0.4\%)$\\
L3 jet-charge (1998)~\cite{jetqL3} & $0.0948 \pm 0.0101 \pm 0.0056$ & $10.8\%$ & $5.9\%$& $4.1\%\,(0.4\%)$\\ 
OPAL jet-charge (2002)~\cite{jetqOPAL}& $0.0994 \pm 0.0034 \pm 0.0018 $& $3.7\%$ & $1.8\%$& $1.5\%\,(0.3\%)$\\\hline
\end{tabular}
\end{center}
\end{table}
At future high-luminosity $\epem$ machines, such as the FCC-ee with $10^5$ times more data collected
at the Z pole than at LEP~\cite{FCCee}, statistical uncertainties will be totally negligible, and 
the latter QCD effects will dominate the systematics of the $\AFBb$ measurement. 


\section{Simulation of the LEP $b$-quark forward-backward asymmetry measurements}
\label{sec:}

The eight original LEP measurements of $\AFBb$ 
have been implemented in a MC event simulation based on \pythia~8.226~\cite{pythia8} with seven 
different parton-shower and hadronization tunes, as well as based on two alternative (dipole antenna) shower
approaches from \pythia~8.210 combined with \vincia~1.1 and 2.2 (with uncertainties given by
12 variations of the \vincia\ parameter set)~\cite{vincia}. Ten million $\epem\to Z(\bbbar)$ events are thereby generated
at $\sqrts = 92.4$~GeV with QED radiation on, 
and analysed as done in the original experiments. The whole MC setup effectively corresponds 
to nine different modelings of the underlying QCD effects (bottom- and charm-quark gluon radiation and 
fragmentation functions, and $B,D$ semileptonic decays). Tune-7 and \vincia~2.2 include proton-proton 
data whereas all other models are based on LEP data alone.
For all analyses, the $b$-jets are first reconstructed with the JADE algorithm 
from the list of final-state particles and the thrust axis of the event is computed as a proxy of the $\bbbar$ 
direction. Each original $y_{\rm cut}$ and $M_{\rm jet}$ jet selection criteria, and 
(transverse) momenta ($p_T$) $p$ cuts on the final electron and muons, are applied. On the one hand, 
the lepton-based analyses determine the $b$-quark charge from that of the hardest charged lepton in the
event, and then extract $\AFBobs$ by fitting the corresponding distribution of polar angles $\theta$ between 
the $e^-$ and the thrust axis, $dN/d\cos\theta = 3/8\,\;[1+\cos^2\theta+8/3\,\AFBobs(1-2\chi_B)\cos\theta]$, where
$\chi_B\approx 0.12$ is the $B^0\overline{B^0}$ effective mixing parameter.
On the other, in the jet-charge-based analyses, $b,\bar{b}$-quarks are identified via their
measured jet charge $Q_{\rm jet}=\sum p_L^\kappa\,Q/\sum p_L^\kappa$ (where $p_L$
is the longitudinal momentum of the final-state particles, with charge $Q$, with respect
to the thrust axis, and the power $\kappa$ varies between 0.4 and 0.6), and $\AFBobs$ is derived 
by fitting the distribution
$\mean{Q_F-Q_B}/\mean{Q_b-Q_{\bar{b}}}=8/3\,\AFBobs(1+C)\cos\theta/(1+\cos^2\theta)$,
where $Q_F\;(Q_B)$ are the jet charges in the forward (backward) hemisphere, and the
$C$ factor is a $\sim$3.5\% correction for missing higher-order QCD terms and for the 
difference between the thrust axis and the $b$-quark direction~\cite{ALEPH:2005ab,Abbaneo:1998xt}. 


\section{Results and conclusions}
\label{sec:}

\begin{figure}[htpb!]
\includegraphics[width=0.47\linewidth,height=4.8cm]{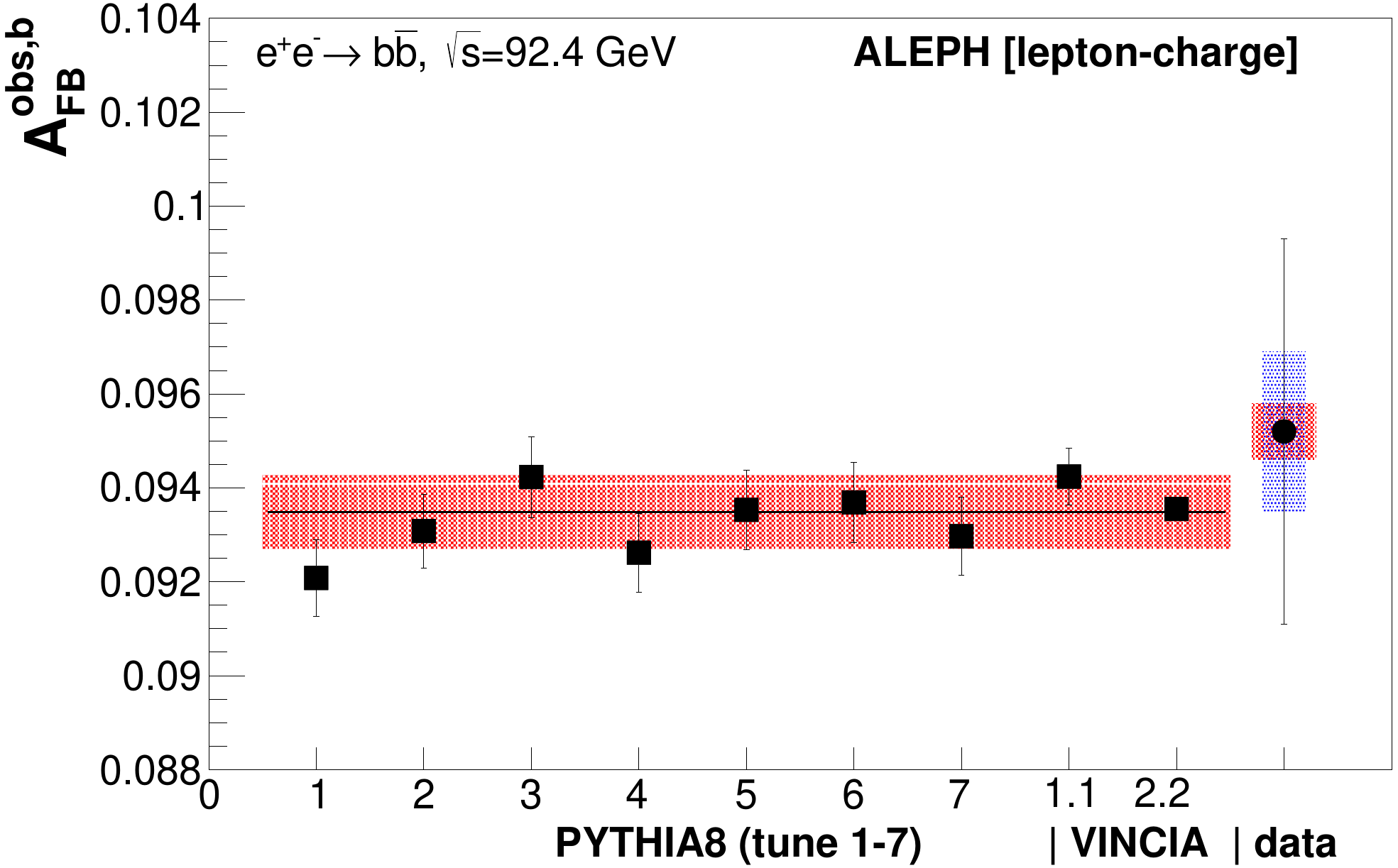}
\includegraphics[width=0.47\linewidth,height=4.8cm]{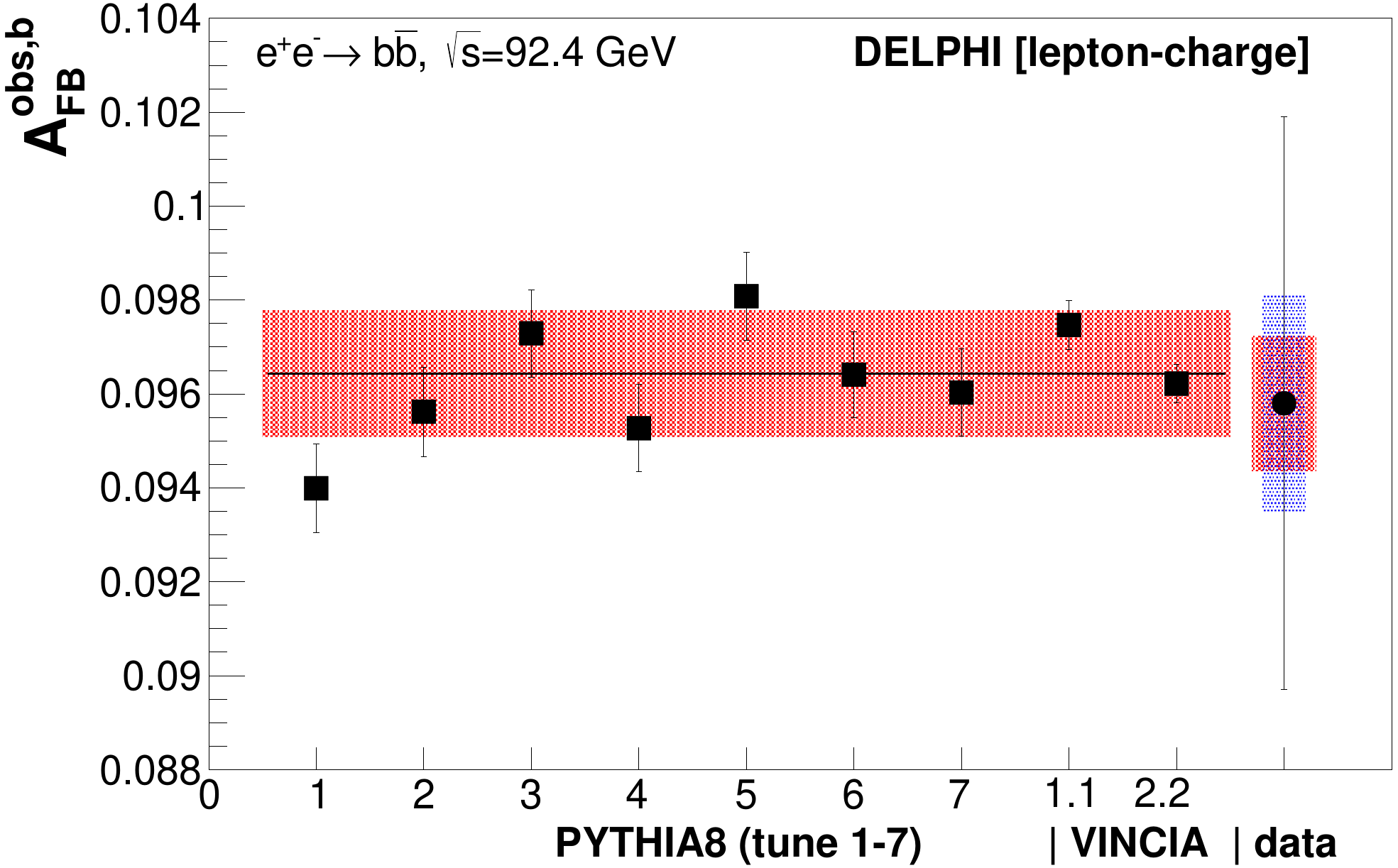}\\
\includegraphics[width=0.47\linewidth,height=4.8cm]{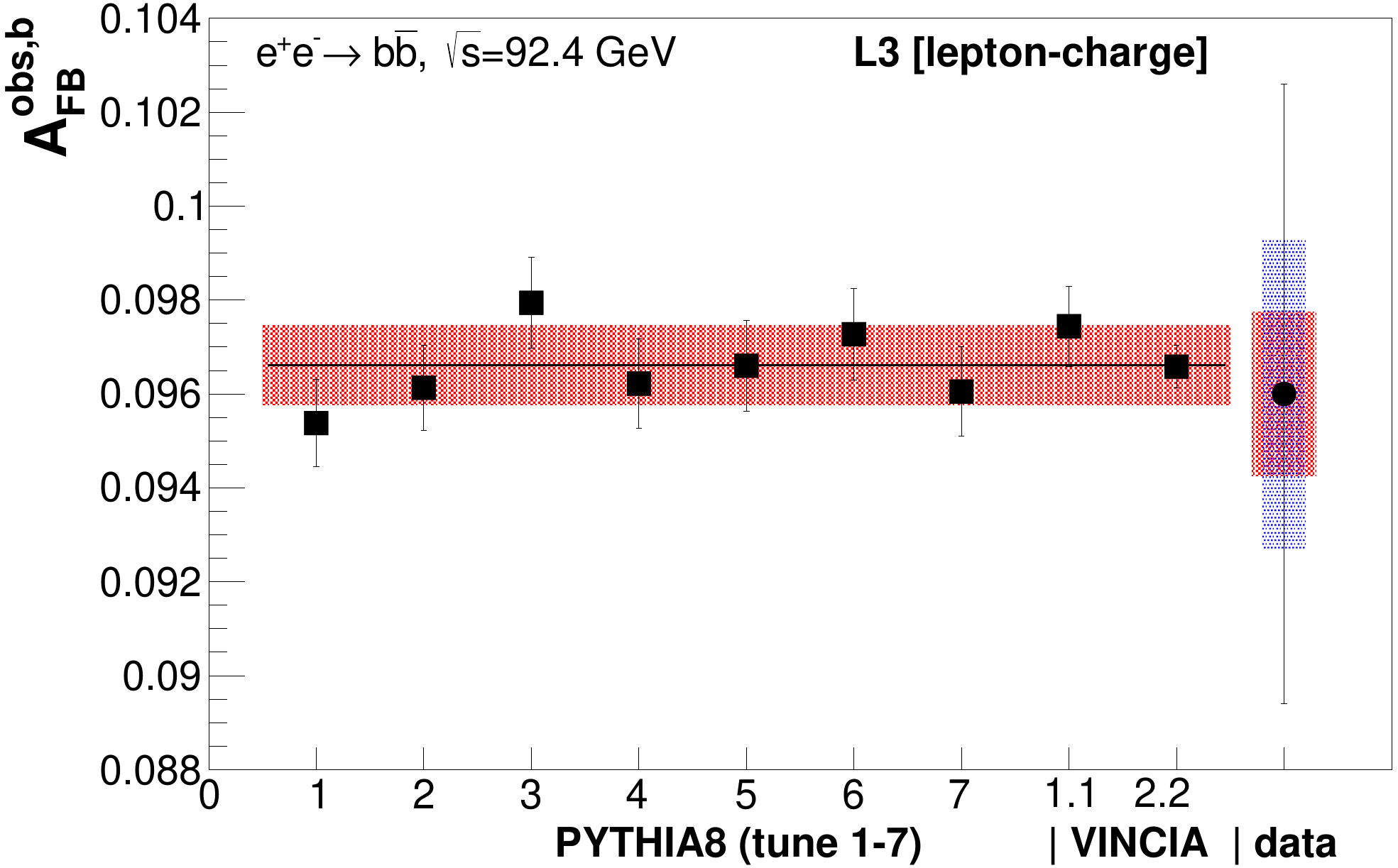}
\includegraphics[width=0.47\linewidth,height=4.8cm]{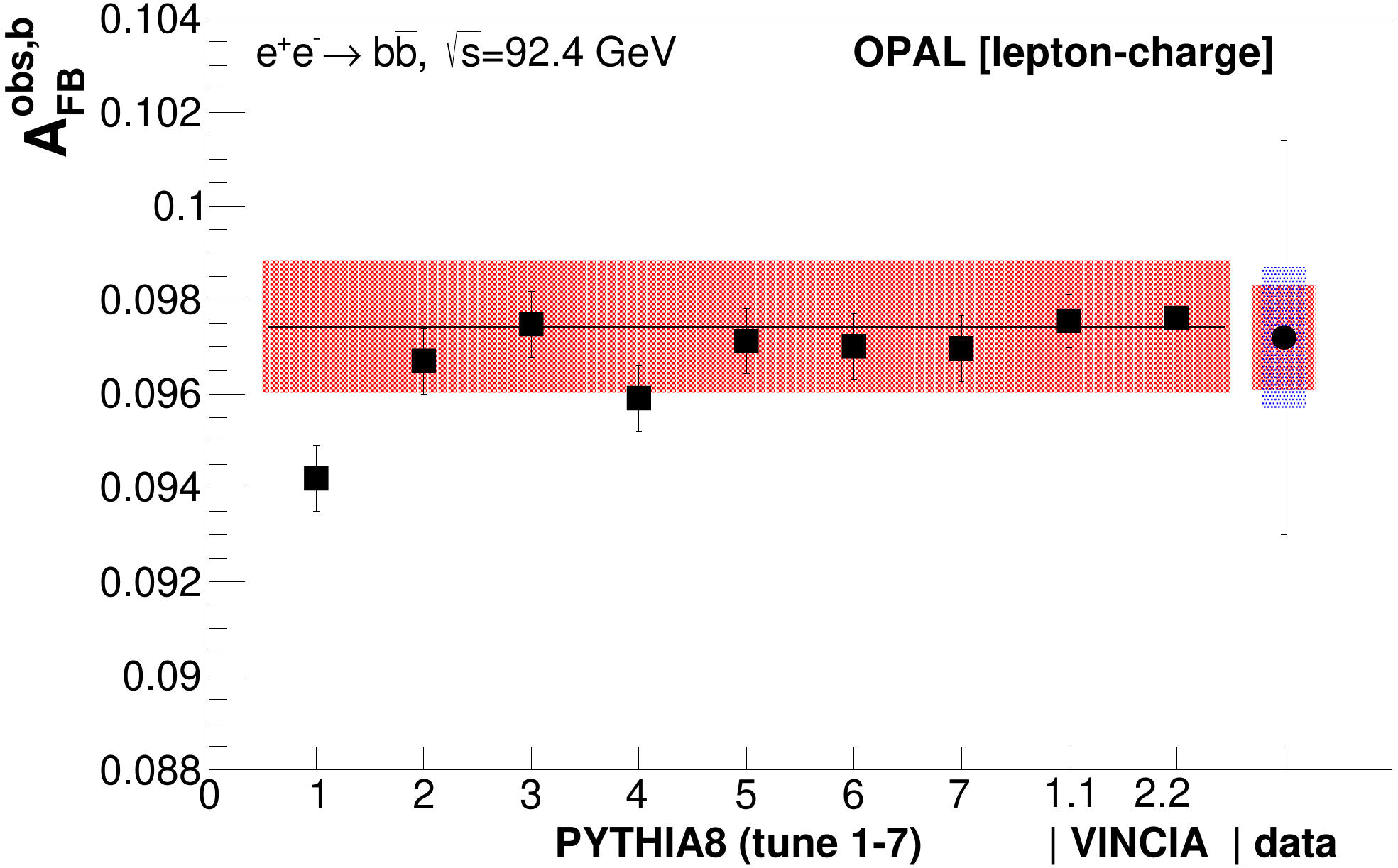}
\caption[]{$b$-quark forward-backward asymmetry 
extracted from lepton-charge
analyses of $\epem\to\bbbar$ simulations based on seven \pythia~8 and 
two \pythia~8+\vincia\ tunes 
(squares with red band), compared to the corresponding
experimental results  (rightmost data point, with QCD, in red, and uncorrelated, in blue, systematic uncertainty bands) 
measured by ALEPH (top left)~\cite{leptonALEPH}, DELPHI 
(top right)~\cite{leptonDELPHI95}, L3 (bottom left)~\cite{leptonL392},
and OPAL (bottom right)~\cite{leptonOPAL}.} 
\label{fig:lepton_AFBobs_vs_MC}
\end{figure}

\begin{figure}[htpb!]
\includegraphics[width=0.47\linewidth,height=4.8cm]{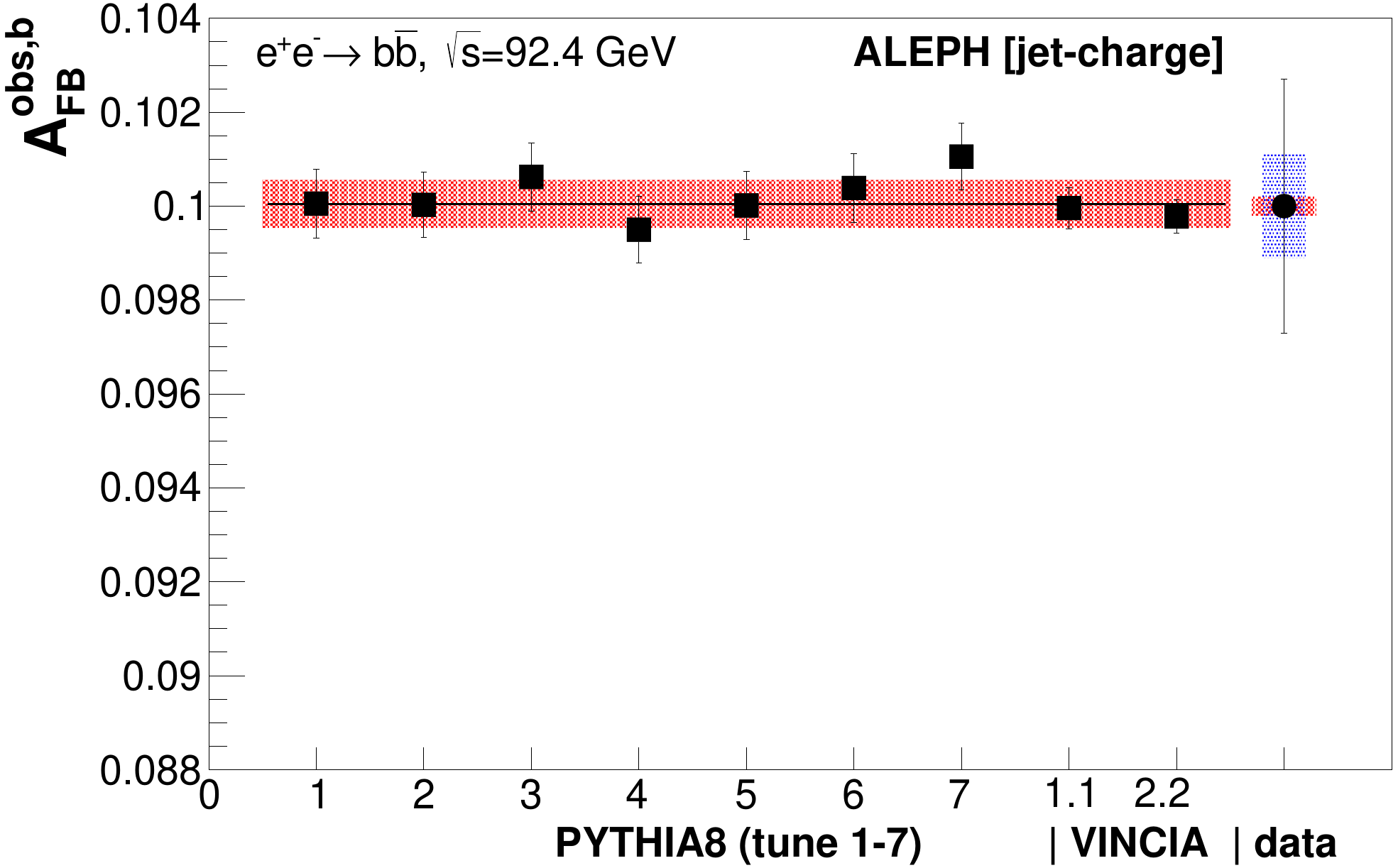}
\includegraphics[width=0.47\linewidth,height=4.8cm]{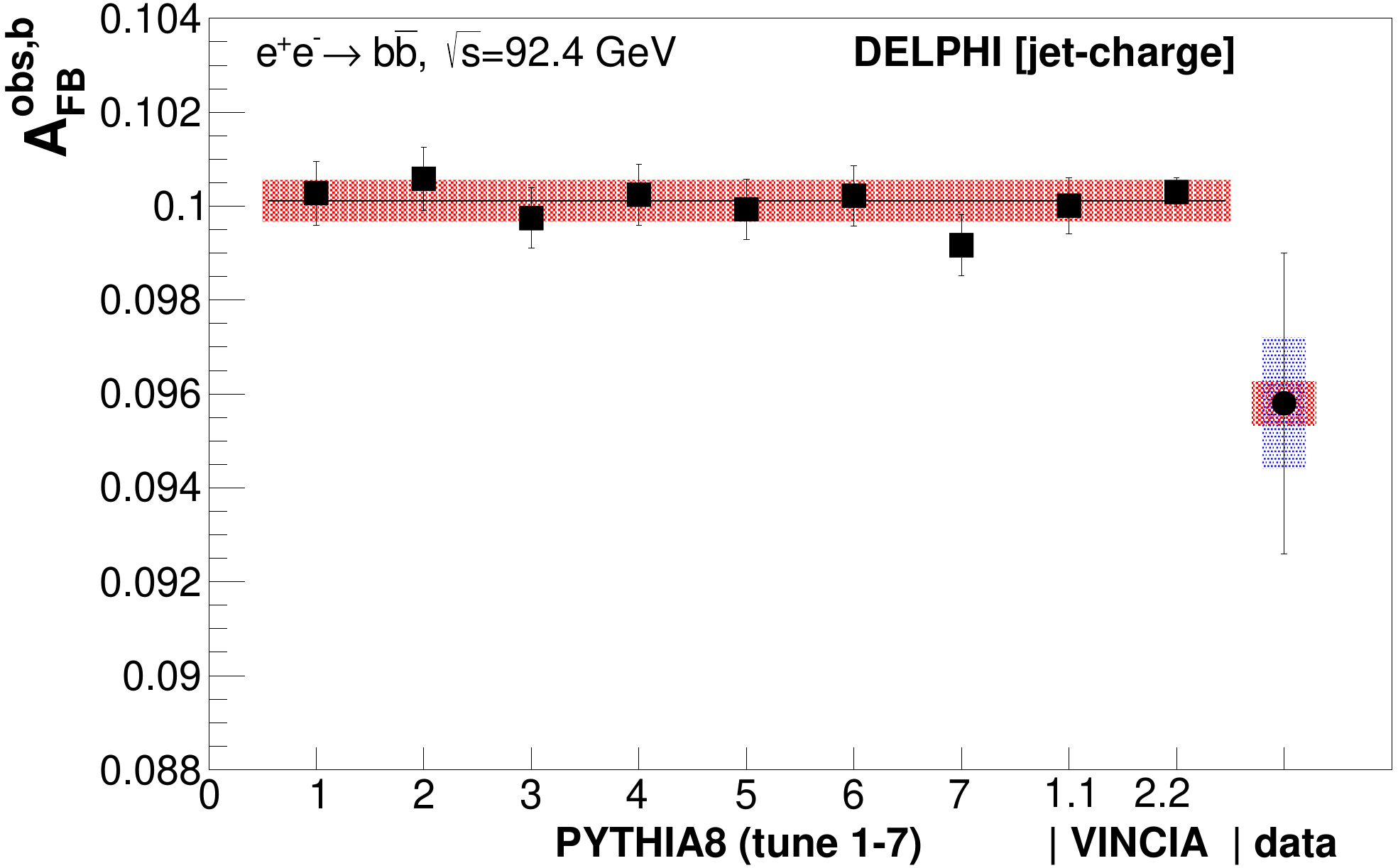}\\
\includegraphics[width=0.47\linewidth,height=4.8cm]{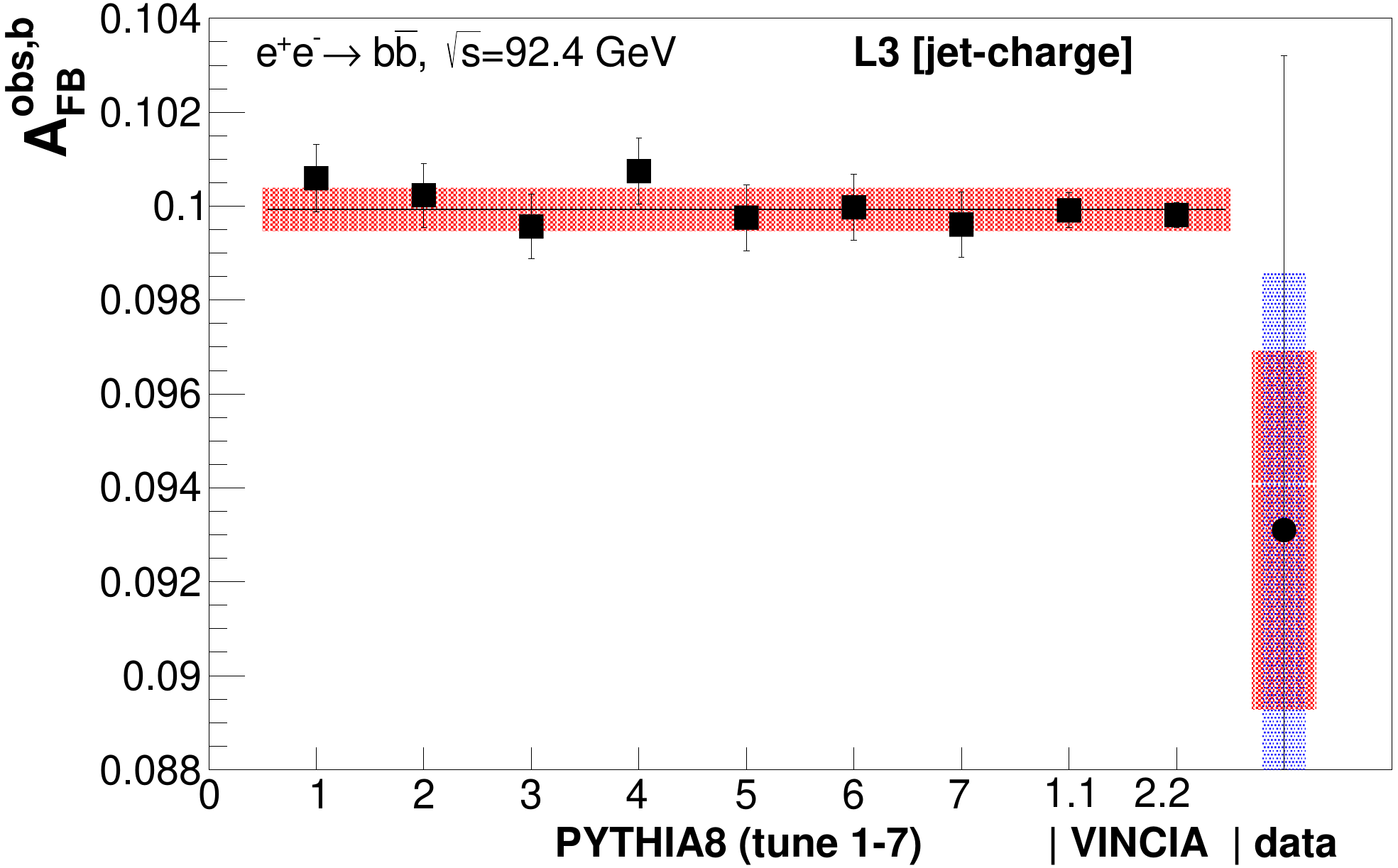}
\includegraphics[width=0.47\linewidth,height=4.8cm]{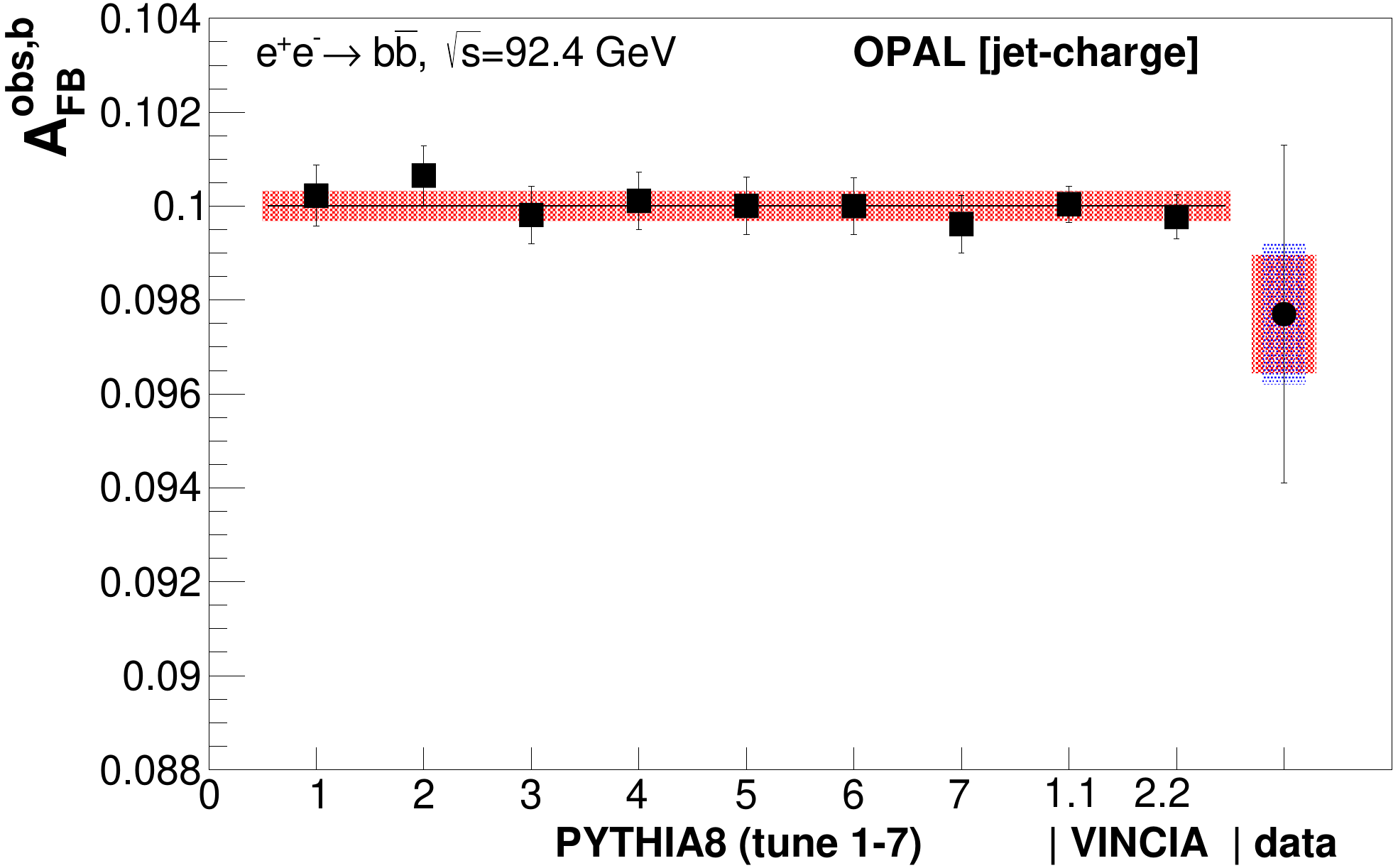}
\caption[]{$b$-quark forward-backward asymmetry 
extracted from jet-charge
analyses of $\epem\to\bbbar$ simulations based on seven \pythia~8 and 
two \pythia~8+\vincia\ tunes 
(squares with red band), compared to the corresponding
experimental results (rightmost data point, with QCD, in red, and uncorrelated, in blue, systematic uncertainty bands) 
measured by ALEPH (top left)~\cite{jetqALEPH}, DELPHI 
(top right)~\cite{jetqDELPHI}, L3 (bottom left)~\cite{jetqL3},
and OPAL (bottom right)~\cite{jetqOPAL}.} 
\label{fig:jet_AFBobs_vs_MC}
\end{figure}

Through the procedure describe above, we extract 9 different MC values of $\AFBobs$ 
for each one of the eight experimental setups, which we compare among themselves and against 
the experimental data in Fig.~\ref{fig:lepton_AFBobs_vs_MC} and~\ref{fig:jet_AFBobs_vs_MC}
for lepton- and jet-charge analyses. 
The central $\AFBobs$ values plotted differ slightly 
from the $\AFBb$ values quoted in Table~\ref{tab:AFBb}, since the latter are obtained correcting for radiative 
effects, $\gamma$ exchange, Z-$\gamma$ interference, and shifted to the pole $m_Z = 91.187$~GeV mass. 
The first (leftmost) MC point corresponds to the
\pythia~8 tune-1 result obtained with the 1990 \jetset\ parameter set, very similar to the
one used to obtain the original LEP QCD 
uncertainties~\cite{Abbaneo:1998xt}. 
The red band around the MC points is the standard deviation of the predictions, 
which we take as indicative of the associated QCD systematic uncertainty for each measurement. It amounts to
about 1\% (0.4\%) for the lepton (jet) charge-based measurements, and is found to be overall slightly smaller but
still fully consistent with the original QCD uncertainties (last column of Table~\ref{tab:AFBb}).
Using the updated QCD systematics, we obtain~\cite{DdEYan} a new weighted-average $b$-quark 
forward-backward asymmetry, $\AFBb = 0.0996\pm0.0016$, very similar to the current one.

\end{document}